\documentclass[aps,prb,twocolumn,showpacs]{revtex4}

\usepackage{graphicx}
\usepackage{dcolumn}
\usepackage{bm}
\usepackage[normalem]{ulem}

\begin{document}

\title{$^{\bf 4}$He adsorbed inside (10,10)  single walled carbon nanotubes}  

\author{M.C.Gordillo}
\affiliation{Departamento de Sistemas F\'{\i}sicos, Qu\'{\i}micos 
y Naturales. Facultad de Ciencias Experimentales. Universidad Pablo de Olavide. 
Carretera de Utrera, km 1. 41013 Sevilla. Spain.}
\author{J. Boronat and J. Casulleras}
\affiliation{Departament de F\'\i sica i Enginyeria Nuclear,
Universitat Polit\`ecnica de Catalunya,
B4-B5 Campus Nord. 08034 Barcelona. Spain.}

\date{\today}

\begin{abstract}
Diffusion Monte Carlo calculations on the adsorption of $^4$He in open-ended
single walled (10,10) nanotubes are presented. We  have found a 
first order phase transition separating 
a low density liquid phase in which  all $^4$He  atoms are adsorbed 
close to the 
tube wall and a high density arrangement characterized by two helium 
concentric layers. 
The energy correction  
due to the presence of neighboring tubes in a bundle has also been
calculated, finding it negligible in the density range considered.  
\end{abstract}

\pacs{68.90.+g,05.30.-d}

\maketitle

One of the most interesting capabilities of  carbon nanotubes and their
bundles is their purported use as reservoirs, in particular for the 
lightest quantum fluids ($^4$He, H$_2$ and
Ne)~\cite{Dillon,talapatra,teizer, vilchesjltp1,vilchesjltp2,glyde}.   The
most obvious  adsorption places are the outer part of the bundles, formed
by the depressions known as grooves between every two tubes, and the rest
of the exposed external surfaces.  These adsorption places have 
already been experimentally documented~\cite{glyde,teizer,
lasjaunias,vilchespa}. One can think also  of the interchannels located
every three -- or more if the  arrangement  is irregular --  tubes in the
inner part of the bundle as  accessible sites,  but this latter possibility
seems  uncertain since any single vacancy in the carbon structure of the
nanotubes could block those very thin interstices~\cite{prl06}.  

The other possible adsorption site is the inner part of the tubes
themselves. This would imply to remove the ending caps of the nanotubes and
thus  allowing the gases to  enter. To our knowledge, there is no
experimental information for this  last possibility in the case of $^4$He,
the majority of theoretical studies being related to the limit of infinite
dilution~\cite{coleRMP} or to  quasi one-dimensional narrow
environments~\cite{inter,prbr,kro}, even  though there are some
exceptions~\cite{collo,reatto,reattolow}. In the present work, we present a
study of $^4$He adsorbed in a (10,10)  single walled carbon nanotube,  one
of the most common types. We have also estimated the effect that
neighboring tubes have in the adsorption  energies of $^4$He inside a
single tube. That  can help to make a good estimation of the  real
adsorption capabilities of nanotube bundles.  

We have carried out a microscopic study of the system at zero
temperature using the Diffusion Monte Carlo (DMC) method.
DMC is a stochastic technique for solving the many-body Schr\"odinger equation,
providing for boson systems exact results within some
statistical uncertainties~\cite{boro94}.
The carbon atoms in the nanotubes have been explicitly considered,
meaning that the carbon nanotube-helium potential was obtained
by summing up all the C-He pair interactions (taken 
from Ref. \onlinecite{cole}), with the carbon atoms in fixed positions. This
is not the most common setup, since most of the previous studies were 
carried out
considering a smoothed helium-tube interaction (an exception being
Ref. \onlinecite{collo}). The helium-helium potential is the HFD-B(HE)
Aziz potential~\cite{aziz}. For the liquid phase simulations (see below),
we have used
$N=150$ $^4$He atoms, and to run calculations at different densities we have
changed the length of the simulation cell between 1537.5 and 34.44 \AA,
these values being in all cases a multiple of the elementary carbon cell of
length 2.46 \AA, characteristic of an armchair nanotube. 
When a solid-liquid system was considered, we fixed the
number of atoms belonging to the solid phase, and varied the number of atoms
in the liquid one. Periodic boundary conditions in the transverse 
direction ($z$ in
cylindrical coordinates) have always been assumed.

The trial wave function needed in the method for guiding the diffusion 
process was chosen to be of the form: 
\begin{equation}
\Psi({\bf r}_1,\ldots,{\bf r}_N) = \prod_{i<j}^{N} f_{\rm{He}-\rm{He}} (r_{ij})
\prod_{i,j=1}^{N,N_c} f_{\rm{He}-\rm{C}} (r_{ij}) \prod_{i=1}^{N} \Phi(r_i)
\ ,
\label{trial}
\end{equation} 
with $N$ and $N_c$ the number of $^4$He and C atoms, respectively, and
$r_i$ the radial distance of particle $i$ to the center of the nanotube.
The first two terms of $\Psi$ in Eq. (\ref{trial})
correspond to two-body correlation factors which account for dynamical
correlations induced by the He-He and He-C potentials. Both
$f_{\rm{He}-\rm{He}}(r)$ and $f_{\rm{He}-\rm{C}}(r)$ are of McMillan type,
$f(r)=\exp[-0.5 (b/r)^5]$. The value for the parameters $b_{\rm{He}-\rm{He}}$ 
has been taken form Ref. \onlinecite{boro94}. The value of 
$b_{\rm{C}-\rm{He}}$ varied from the phases with one helium layer 
($b_{\rm{He}-\rm{He}}$ = 0) to the ones with two helium layers 
($b_{\rm{He}-\rm{He}}$ = 2.3 \AA, see below). In both cases, $b$ was obtained by means 
of separate Variational Monte Carlo (VMC) calculations and found 
to be independent on the helium density. 

The variance of the calculation is reduced by introducing also in the trial
wave function (\ref{trial}) a one-body 
term $\Phi(r_i)$ which accounts for the mean
interaction between a single He atom and the nanotube wall. At small
densities, we have taken for $\Phi(r_i)$ the solution of the Schr\"odinger
equation for a single $^4$He atom in the cylindrically averaged potential of 
Ref. \onlinecite{cole} for a tube of the same diameter than a (10,10) one.
This function has a single peak located roughly in the minimum of
the considered helium-tube potential.
When the density increases one can see the emergence of a second peak in
the density profile obtained from the DMC simulations. This peak 
is located nearer to the center of the nanotube than the previous one.
Thus, in the 
high density regime we verified that the variance is further reduced by
changing the low-density one-body term to another one 
which contains this closer-to-the-center peak in addition to the one
near the carbon wall.  
In particular, we have used for $\Phi(r_i)$ the square root of the 
radial density profile
obtained in a DMC calculation at density  $\rho=0.028$ \AA$^{-3}$. This
trial wave function has two maxima and a reduced density (but not zero)
between them.
In principle, this profile with the double peak would correspond to a liquid
phase since no localization on fixed sites of any of the helium atoms was 
imposed. 

When the density increases even more the layer closer to the wall can
crystallize. In order to characterize this possible solid layer we have also
studied the system by introducing Nosanow-Jastrow terms describing a solid
order around the cylinder,
\begin{equation}
\Psi_{\rm s} ({\bf r}_1,\ldots,{\bf r}_N) = \prod_{i=1}^{N_{\rm s}} 
g(|{\bf r}_i - {\bf R}_i|) \  \Psi ({\bf r}_1,\ldots,{\bf r}_N)
\ ,
\label{trialsol}
\end{equation} 
with $\Psi$ the trial wave function of the liquid phase with a double
peak (\ref{trial}) and
$g(r)$ a gaussian factor localizing a given set of particles around 
predefined lattice sites ${\bf R}_i$. The lattice chosen for the simulation
corresponds to a triangular-like planar lattice rolled up to form a
cylinder concentric to the carbon nanotube. A VMC optimization process
was performed first by using a fixed number of atoms (150) for different lengths
of the simulation cell, and varying also the number of atoms located in 
the first and second helium layers. We then compared the simulation results
to those without the gaussian constrains at the same density. 
Those set of  optimizations leaded us to conclude that the minimum energy  
configurations in the density range studied correspond to a lattice 
formed by slices of eight atoms 
that approximately share the same $z$ coordinate (that of the tube axis) 
and are separated by a distance of 2.46 \AA\ in the $z$ direction. This layer is 
located at a distance of 2.8 \AA\ from the carbon wall. Any other 2D
solid phases, including registered ones we found to produce configurations
of higher energy. 
In the final simulations, we have used 14 nanotube cells that, with
this optimized structure, represent a solid layer composed by 112 $^4$He atoms. 
The rest of the helium, up to 150 atoms, was considered to be in the 
second layer and was disposed without gaussian restrictions. 
The free parameter of the gaussian $g(r)=\exp (-\alpha r^2)$ has been
optimized ($\alpha$=2.30 \AA$^{-2}$) showing negligible dependence on the
density in the range studied. A check was also made to see the influence of
the corrugation imposed by the carbon structure 
in the energy of the solid layer. To do so, we located the solid lattice 
in different positions relative to the underlying structure. The 
energies obtained were similar within the error bars, what indicates that
the solid layer is a two-dimensional incommensurate solid. 
 
The results for the total set of simulations of $^4$He inside a single 
(10,10) tube are displayed in Fig. 1.
The densities were calculated by straightforwardly dividing the
number of atoms by the nanotube volume, considering a 
radius of 6.8 \AA\ in all cases. Full circles are the results for a one-layer
He phase, open circles correspond to the two-layer liquid arrangement, and 
the full squares indicate the energies obtained for the phase with a solid
layer closer to the carbon wall. The main conclusion one can draw from
this figure is that there is a minimum of energy of the one-shell phase;
corresponding to zero pressure and that can be found from a 
third order polynomial fit to be at 0.0086 \AA$^{-3}$ with a corresponding
binding energy of 187.18 $\pm$ 0.02 K. The same binding energy for the
infinite dilution limit is 185.01 $\pm$ 0.02 K. That is 
bigger than the experimental value for graphite, 143 $\pm$ 2 K \cite{elgin},
and smaller than the corresponding values for a
(5,5) tube (429.984 and 429.966 K at zero pressure and for infinite
dilution, respectively). This simply means that an increasing in the tube
radius implies a lowering of the binding energy to approximate the one
in the flat surface of graphite. What is more important, the stable
phase at zero pressure and 0 K is a $liquid$.  
In this respect, the situation is different than in
the case of graphite \cite{manousakis} and for the cylindrical system
considered in Ref. \onlinecite{reatto}. 

\begin{figure}
\setlength\fboxsep{0pt} \centering
\scalebox{1.00}{\includegraphics[width=1.0\linewidth]{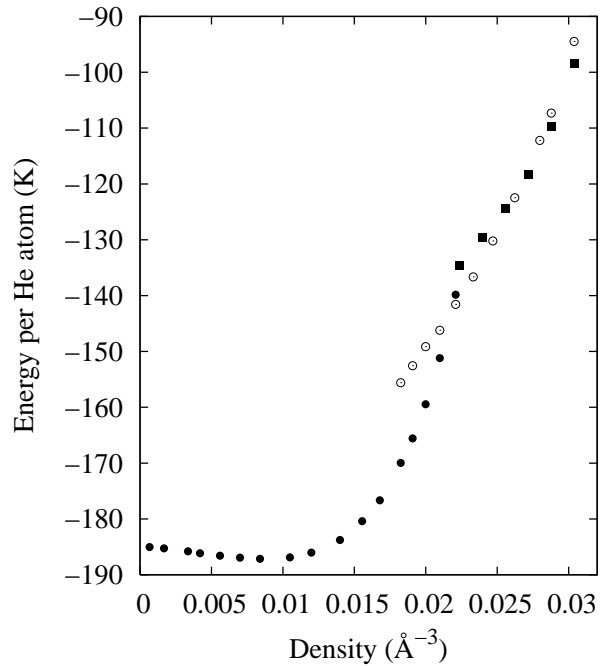}}
\caption{Energy versus density for different elections of the trial 
function: One layer liquid (full circles); 
(squares); two layer liquid (circles); a close to the wall solid
(full squares). 
}
\label{fig1}
\end{figure}

The stability of the rest of the phases can be deduced with the help
of Fig. 2. This is a Maxwell construction for the density range in 
which we could have a two-layer liquid or a two-layer solid+liquid.
Full lines are third-grade polynomial fits to the corresponding 
one layer liquid phase, and the two-layer solid+liquid arrangement, while
circles are the results for the two-layer liquid. The x axis corresponds 
to the inverse of the density as defined in the previous figure.
The dashed line is a double tangent construction between the one-layer
liquid and the phase including a solid. In a conservative frame, 
this would indicate a transition between a low density phase of 
2 10$^{-2}$ \AA$^{-3}$ and
a one including a solid,  of 2.7 10$^{-2}$ \AA$^{-3}$. The transition 
pressure would be of 430 atm. However, another alternative explanation
can be deduced from the data, and it is a double transition between
a single layer liquid and a double layer liquid, followed by another
one between this last dense liquid and the solid and liquid phase. 
The corresponding coexistence densities are the same for the two phases already
mentioned, but we will have small a window with a two-layer 
liquid at around 2.5 10$^{-2}$ \AA$^{-3}$.
The one shell liquid- two shell liquid transition would be then at 420 atm. 
Unfortunately,
the available data (and shown in Table I) will not allow us to discriminate
between both scenarios. In any case, it is clear that an increasing in the
density will eventually create a solid close to the wall.   

\begin{figure}
\setlength\fboxsep{0pt} \centering
\scalebox{1.00}{\includegraphics[width=0.8\linewidth]{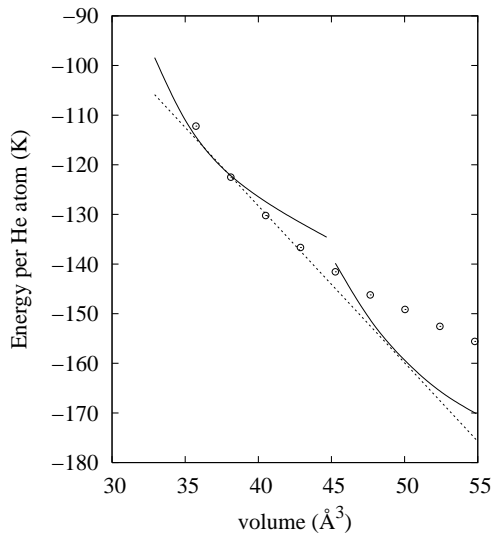}}
\caption{Maxwell construction to determine the stability range of
the phases involved in the modeling of the system.
Full line, one layer liquid and solid+liquid phases ; 
Circles, two layer liquid phase. 
}
\label{fig2}
\end{figure}

To complete the characterization of the phases, we displayed the radial 
profiles for different helium densities in Fig. 3. 
The dashed line represents the case of zero pressure (0.0086 \AA$^{-3}$),
while the dotted line displays the situation in the low
density limit of the transition region. One can see that in both cases
the system is formed by a single cylindrical helium shell, located at
around 4 \AA\ from the center of the tube. The main difference 
is the increasing in the helium density that can be inferred 
from the growth in the corresponding peak. The long-dashed line indicates
the situation in the upper density limit of the coexistence zone 
for the case of the solid+liquid phase. We observe
a second peak located closer to the center of the tube, and another one
corresponding to the two-dimensional solid,
roughly in the same position than in the single layer phase. 
Finally, the full line represents the solid+liquid phase at the highest
density simulated. There, the first peak is virtually identical to the 
one at the transition, as befits to a solid of unchanging density, going
the additional helium to fill the second layer. 

\begin{figure}
\setlength\fboxsep{0pt} \centering
\scalebox{1.00}{\includegraphics[width=0.8\linewidth]{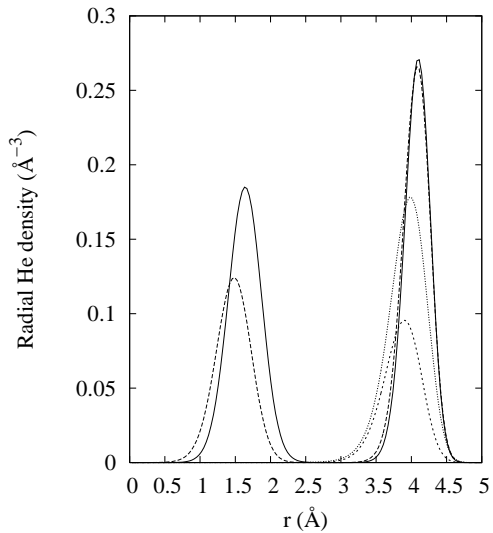}}
\caption{Radial density profiles at zero pressure (dashed line), 
at the lower and upper limits of the coexistence regions (dotted and 
long dashed line, respectively). The solid line represents the case
for the solid+liquid phase at the highest density simulated. 
}
\label{fig3}
\end{figure}

Until now, we considered only the case of an isolated carbon nanotube, 
but this is not the most common situation, since those cylinders tend to
group to form what is termed a bundle. There, each tube is ideally surrounded
by six others, forming a structure whose section is similar to a triangular
2D lattice with a minimum distance between centers of cylinders of 17 \AA\
\cite{Tersoff}.  Taking that into account, we estimated
the influence that the surrounding tubes have in the energy per helium atom
within a mean field scheme by means of the following expression: 
\begin{eqnarray}
\lefteqn{E_{correction}  =  6 \int_{x'} \int_{y'} d(x',y') dx' dy'}  \\
& & \times \int_0^\infty \int_x \int_y R(x,y,z) V(x,y,z,x',y') dx dy
dz
\nonumber
\end{eqnarray}
where $d(x',y')$ represents the normalized probability of finding an $^4$He
atom at coordinates $x'$ and $y'$ for any $z$ position in the first tube.   
$R(x,y,z)$ is the corresponding radial density function as the ones depicted
in Fig. 3, and depends on the location of helium atoms in the second
tube ($x,y,z$). Finally, $V(x,y,z,x',y')$ is the helium-helium Aziz potential
used in our simulations \cite{aziz}. The result of the integral is multiplied
by six to take into account the number of tubes surrounding a particular one
at the same distance. A check
of the accuracy of this kind of mean field estimation has already been done 
for a closely related system \cite{isotopos} and found a very good approximation
to the results of a full Monte Carlo calculation. 

\begin{table}
\caption{Energies per $^4$He atom in K at several densities.
}
\begin{tabular}{ccc} \hline
Density (\AA$^{-3}$)   & (10,10) tube &  $E_{correction}$ \\ \hline
Low density phase \\ \hline
6.7 10$^{-4}$  &  -185.02(4)  &  -0.003 \\
1.7 10$^{-3}$  &  -185.27(3)  &  -0.006 \\
3.4 10$^{-2}$  &  -185.81(3)  &  -0.013 \\
4.2 10$^{-3}$  &  -186.14(4)  &  -0.016 \\
5.6 10$^{-3}$  &  -186.58(3)  &  -0.022 \\
7.0 10$^{-3}$  &  -186.93(4)  &  -0.027 \\
8.4 10$^{-3}$  &  -187.14(3)  &  -0.033 \\
1.05 10$^{-2}$  &  -186.89(3)  &  -0.041 \\
1.2 10$^{-2}$  &  -186.02(4)  &  -0.047 \\
1.4 10$^{-2}$  &  -183.76(4)  &  -0.055 \\
1.6 10$^{-2}$  &  -180.40(4)  &  -0.062 \\
1.7 10$^{-2}$  &  -176.66(4)  &  -0.067 \\
1.8 10$^{-2}$  &  -169.97(4)  &  -0.073 \\
1.9 10$^{-2}$  &  -165.59(6)  &  -0.077 \\
2.0 10$^{-2}$  &  -159.45(7)  &  -0.081 \\ \hline
Two layer liquid density phase \\ \hline
1.8 10$^{-2}$ &  -155.60(9) &  -0.069 \\
1.9 10$^{-2}$ &  -152.56(9) &  -0.072 \\
2.0 10$^{-2}$ &  -149.15(7) &  -0.075 \\
2.1 10$^{-2}$ &  -146.22(9) &  -0.079 \\
2.2 10$^{-2}$ &  -141.58(6) &  -0.083 \\
2.3 10$^{-2}$ &  -136.66(6) &  -0.087 \\
2.5 10$^{-2}$ &  -130.22(8) &  -0.092 \\
2.6 10$^{-2}$ &  -122.49(8) &  -0.099 \\
2.8 10$^{-2}$ &  -112.2(1)  &  -0.103 \\ \hline
Two layer liquid+solid density phase \\ \hline
2.2 10$^{-2}$ & -134.58(5) &  -0.092 \\
2.4 10$^{-2}$ & -129.63(5) &  -0.095 \\
2.6 10$^{-2}$ & -124.50(5) &  -0.099 \\
2.7 10$^{-2}$ & -118.30(5) &  -0.103 \\
2.9 10$^{-2}$ & -109.64(6) &  -0.107  \\
3.0 10$^{-2}$ &  -98.41(9) &  -0.111 \\
\end{tabular}
\end{table} 
The results of this energy correction are given in Table I. 
One thing is immediately apparent, and it is that the mean field term 
increases with density, but in all cases
is negligible and of the order of the error bars of the DMC calculation.
That means that the binding energy of $^4$He in a bundle is essentially the
same that the one for a single tube.

Summarizing, we have performed a full DMC calculation on the subject of $^4$He 
adsorbed in carbon nanotube bundles. We have found a low-density liquid phase
characterized by a single helium layer close to the nanotube wall. When the
density increases there is a phase transition to a two-layer arrangement
that will eventually solidify with further increase of the number of adsorbed
helium atoms.  We also saw that 
the influence of the neighbouring tubes in a bundle in the properties of
a given one is negligible. Experimental signatures of the phases predicted
in this work could be obtained by means of low temperature specific heat
measurements~\cite{lasjaunias} and neutron diffraction~\cite{glyde}.  

\begin{acknowledgments}
We acknowledge partial financial support from 
the Spanish Ministry of Education and Science (MEC)  
(grants FIS2006-02356-FEDER and FIS2005-04181), Junta de Andalucia
(group FQM-205) and Generalitat de Catalunya (grant 2005GR-00779).
\end{acknowledgments}

\end{document}